\documentclass[prapplied,
superscriptaddress,showpacs,amsmath,amssymb,  floatfix, twocolumn, longbibliography]{revtex4-2}

\usepackage{graphicx}
\usepackage{dcolumn}
\usepackage{bm}
\usepackage{xcolor}
\usepackage{ulem}
\usepackage[utf8]{inputenc}

\usepackage{siunitx}
 \DeclareSIUnit\bar{bar}

\usepackage{hyperref}
\usepackage{booktabs}

\usepackage{tabularx}
\usepackage{multirow}
\usepackage{rotating}
\newcommand\tabrotate[1]{\begin{turn}{90}\rlap{#1}\end{turn}}

\newcommand{\ps}[1]{\textcolor{blue}{#1}}

\usepackage{dcolumn} 
\usepackage{bm} 

\DeclareSIUnit{\dB}{dB}

\newcommand{\Omegam}{\Omega_{\textrm{m}}} 
\newcommand{\Gammam}{\Gamma_{\textrm{m}}} 
\newcommand{\Gammaeff}{\Gamma_{\textrm{eff}}} 
\newcommand{\Qm}{Q_{\textrm{m}}} 
\newcommand{\mtot}{m} 
\newcommand{\rp}{R} 
\newcommand{\rhop}{\rho_{\mathrm{p}}} 
\newcommand{\zmean}{\bar{z}} 
\newcommand{\zop}{\hat{z}} 
\newcommand{\nm}{n_{\mathrm{m}}} 
\newcommand{\bgen}{\hat{b}^{\dagger}} 
\newcommand{\bkill}{\hat{b}} 
\newcommand{\xzpf}{\zeta_{\mathrm{zpf}}} 
\newcommand{\nth}{n_{\mathrm{th}}} 
\newcommand{\bz}{b_{\mathrm{z}}} 
\newcommand{\bi}{b_{\mathrm{i}}} 
\newcommand{\Phim}{\Phi_{\mathrm{pul}}} 
\newcommand{\Phiind}{\Phi_{\mathrm{s}}} 
\newcommand{\Ff}{\beta}     
\newcommand{\atech}{\alpha} 
\newcommand{\Gmech}{G} 
\newcommand{\Gmechy}{G_{\mathrm{y}}} 
\newcommand{\gNull}{g_{0}} 
\newcommand{\Hint}{\hat{H}_{\mathrm{int}}} 
\newcommand{\nr}{\bar{n}_{\mathrm{r}}} 
\newcommand{\Simp}{S_{\mathrm{imp}}} 
\newcommand{\Sxximp}{S_{\mathrm{xx}}^{\mathrm{imp}}} 

\newcommand{\etad}{\eta_{\mathrm{d}}} 
\newcommand{\Sin}{S_{\mathrm{in}}} 
\newcommand{\Sout}{S_{\mathrm{out}}} 
\newcommand{\Iext}{I_{\mathrm{ext}}} 
\newcommand{\Vext}{V_{\mathrm{ext}}} 
\newcommand{\Itrap}{I_{\mathrm{trap}}} 
\newcommand{\Ifb}{I_{\mathrm{cb}}} 
\newcommand{\Vfb}{V_{\mathrm{cb}}} 
\newcommand{\Ltot}{\lambda_{\mathrm{tot}}} 
\newcommand{\Lrt}{\lambda_{\mathrm{rt}}} 
\newcommand{\Lc}{\lambda_{\mathrm{c}}} 
\newcommand{\Latt}{\lambda_{\mathrm{att}}} 
\newcommand{\wres}{\omega_{\mathrm{r}}} 
\newcommand{\wrN}{\omega_{0}} 
\newcommand{\widthr}{w_{\mathrm{r}}} 
\newcommand{\sr}{s_{\mathrm{r}}} 
\newcommand{\lr}{l_{\mathrm{r}}} 
\newcommand{\kext}{\kappa_{\mathrm{ext}}} 
\newcommand{\kint}{\kappa_{\mathrm{int}}} 
\newcommand{\ktot}{\kappa_{\mathrm{tot}}} 
\newcommand{\Lr}{L_{\mathrm{r}}} 
\newcommand{\Cr}{C_{\mathrm{r}}} 
\newcommand{\Cext}{C_{\mathrm{ext}}} 
\newcommand{\muN}{\mu_{\mathrm{0}}} 
\newcommand{\eeff}{\epsilon_{\mathrm{eff}}} 
\newcommand{\agen}{\hat{a}^{\dagger}} 
\newcommand{\akill}{\hat{a}} 
\newcommand{\slope}{\textit{f}_{\Phi}} 
\newcommand{\Ws}{W_{\mathrm{s}}} 
\newcommand{\Ds}{D_{\mathrm{s}}} 
\newcommand{\Aeff}{A_{\mathrm{eff}}} 
\newcommand{\Ls}{L_{\mathrm{s}}} 
\newcommand{\AJJ}{A_{\mathrm{JJ}}} 
\newcommand{\Ics}{I_{\mathrm{c}}} 
\newcommand{\Lj}{L_{\mathrm{J}}} 
\newcommand{\betaL}{\beta_{\mathrm{L}}} 
\newcommand{\widthi}{w_{\mathrm{i}}} 
\newcommand{\Si}{s_{\mathrm{i}}} 
\newcommand{\Numberi}{N_{\mathrm{i}}} 
\newcommand{\Wi}{W_{\mathrm{i}}} 
\newcommand{\Lfc}{L_{\mathrm{i}}} 
\newcommand{\Mi}{M} 
\newcommand{\nui}{\nu} 
\newcommand{\DPUL}{D_{\mathrm{pul}}} 
\newcommand{\SPUL}{s_{\mathrm{pul}}} 
\newcommand{\LPUL}{L_{\mathrm{pul}}} 
\newcommand{\Ltw}{L_{\mathrm{t}}} 
\newcommand{\enbw}{\mathrm{ENBW}} 
\newcommand{\Sphiphi}{S_{{\Phi\Phi}}} 
\newcommand{\SVV}{S_{{\mathrm{VV}}}} 
\newcommand{\Sww}{S_{\omega\omega}} 
\newcommand{\Sxx}{S_{\mathrm{xx}}} 
\newcommand{\Syy}{S_{\mathrm{yy}}} 
\newcommand{\SimpC}{S_{\mathrm{imp}}^{\mathrm{co}}} 
\newcommand{\SimpD}{S_{\mathrm{imp}}^{\mathrm{det}}} 
\newcommand{\Sgs}{S_{\mathrm{gs}}} 
\newcommand{\etae}{\eta_{\mathrm{e}}} 
\newcommand{\Cq}{C_{\mathrm{q}}} 
\newcommand{\nmmin}{\nm^{\mathrm{min}}} 
\newcommand{\etacav}{\eta_{\mathrm{d}}^{\mathrm{cav}}} 
\newcommand{\etacryo}{\eta_{\mathrm{d}}^{\mathrm{cold}}} 
\newcommand{\etawarm}{\eta_{\mathrm{d}}^{\mathrm{warm}}} 
\newcommand{\Ttot}{T_{\mathrm{tot}}} 
\newcommand{\nadd}{n_{\mathrm{add}}} 
\newcommand{\nHEMT}{n_{\mathrm{HEMT}}} 
\newcommand{\chim}{\chi_{\mathrm{m}}} 

\newcommand{\magvecpot}{\mathbf{A}(\mathbf{r}, \mathbf{r_0})} 

\newcommand{\kB}{k_{\mathrm{B}}} 
\newcommand{\Tth}{T_{\mathrm{th}}} 

\newcommand{\BI}{b_{\mathrm{trap}}} 

\begin{document}


\title{Remote sensing of a levitated superconductor with a flux-tunable microwave cavity}

\author{Philip Schmidt$^*$}
\email{Philip.Schmidt@oeaw.ac.at}
\affiliation{IQOQI Wien, {\"O}sterreichische Akademie der Wissenschaften, Boltzmanngasse 3, 1090 Wien, Austria}

\author{Remi Claessen$^*$}
\affiliation{Universit{\"a}t Wien, Fakult{\"a}t f{\"u}r Physik, Boltzmanngasse 5, 1090 Wien, Austria}

\author{Gerard Higgins}
\affiliation{IQOQI Wien, {\"O}sterreichische Akademie der Wissenschaften, Boltzmanngasse 3, 1090 Wien, Austria}
\affiliation{Department of Microtechnology and
Nanoscience (MC2), Chalmers University of Technology, Kemiv{\"a}gen 9, 41296 Gothenburg, Sweden}

\author{Joachim Hofer}
\affiliation{IQOQI Wien, {\"O}sterreichische Akademie der Wissenschaften, Boltzmanngasse 3, 1090 Wien, Austria}
\affiliation{Universit{\"a}t Wien, Fakult{\"a}t f{\"u}r Physik, Boltzmanngasse 5, 1090 Wien, Austria}

\author{Jannek J. Hansen}
\affiliation{Universit{\"a}t Wien, Fakult{\"a}t f{\"u}r Physik, Boltzmanngasse 5, 1090 Wien, Austria}

\author{Peter Asenbaum}
\affiliation{IQOQI Wien, {\"O}sterreichische Akademie der Wissenschaften, Boltzmanngasse 3, 1090 Wien, Austria}

\author{Martin Zemlicka}
\affiliation{Institute of Science and Technology Austria, 3400 Klosterneuburg, Austria}

\author{Kevin Uhl}
\affiliation{Physikalisches Institut, Center for Quantum Science (CQ) and LISA$^+$, Universit{\"a}t T{\"u}bingen, 72076 T{\"u}bingen, Germany}

\author{Reinhold Kleiner}
\affiliation{Physikalisches Institut, Center for Quantum Science (CQ) and LISA$^+$, Universit{\"a}t T{\"u}bingen, 72076 T{\"u}bingen, Germany}

\author{Rudolf Gross}
\affiliation{Walther-Mei{\ss}ner-Institut, Bayerische Akademie der Wissenschaften, 85748 Garching, Germany}
\affiliation{School of Natural Sciences, Technische Universit{\"a}t M{\"u}nchen, 85748 Garching, Germany}
\affiliation{Munich Center for Quantum Science and Technology (MCQST), 80799 M{\"u}nchen, Germany}

\author{Hans Huebl}
\affiliation{Walther-Mei{\ss}ner-Institut, Bayerische Akademie der Wissenschaften, 85748 Garching, Germany}
\affiliation{School of Natural Sciences, Technische Universit{\"a}t M{\"u}nchen, 85748 Garching, Germany}
\affiliation{Munich Center for Quantum Science and Technology (MCQST), 80799 M{\"u}nchen, Germany}

\author{Michael Trupke}
\email{Michael.Trupke@oeaw.ac.at}
\affiliation{IQOQI Wien, {\"O}sterreichische Akademie der Wissenschaften, Boltzmanngasse 3, 1090 Wien, Austria}

\author{Markus Aspelmeyer}

\affiliation{IQOQI Wien, {\"O}sterreichische Akademie der Wissenschaften, Boltzmanngasse 3, 1090 Wien, Austria}
\affiliation{Universit{\"a}t Wien, Fakult{\"a}t f{\"u}r Physik, Boltzmanngasse 5, 1090 Wien, Austria}

\date{\today}

\begin{abstract}
We present a cavity-electromechanical system comprising a superconducting quantum interference device which is embedded in a microwave resonator and coupled via a pick-up loop to a $\SI{6}{\micro\gram}$ magnetically-levitated superconducting sphere. The motion of the sphere in the magnetic trap induces a frequency shift in the SQUID-cavity system. We use microwave spectroscopy to characterize the system, and we demonstrate that the electromechanical interaction is tunable. The measured displacement sensitivity of $\SI{e-7}{\metre\per\sqrt{\hertz}}$ defines a path towards ground-state cooling of levitated particles with Planck-scale masses at millikelvin environment temperatures.
\end{abstract}
\maketitle

\section{Introduction}
Superconducting circuits are a cornerstone of quantum technology. They have enabled milestone demonstrations in quantum information processing, quantum-limited amplification, as well as quantum logic-enhanced sensing and control \cite{Wallraff2004,Niemczyk2010,Arute2019,Bal2012,Shanks2013,Danilin2018, Toida2023}.
Superconducting quantum interference devices (SQUIDs) are among the most sensitive detectors of magnetic flux, and have been implemented for the detection of mechanical motion in clamped and levitated micromechanical oscillators \cite{Clarke1989, Clarke2004, Degen2017, Mitchell2020, Etaki2008, vanHeck2023,Vinante2020, Hofer2023, Fuchs2024}. Recently, inductive coupling of mechanical oscillators to microwave circuits with integrated SQUIDs was shown \cite{Rodrigues2019, Schmidt2020, Zoepfl2020, Bera2021, Bothner2022, Luschmann2022, Zoepfl2023}.  Compared to direct-current readout,  microwave spectroscopy of the coupled SQUID-cavity system is advantageous as it provides access to technological advances in circuit quantum electrodynamics. For example, it enables the use of squeezed light in quantum-limited Josephson parametric amplifiers \cite{Teufel2009, Clark2016, Yamamoto2008, Menzel2012, Macklin2015, Fedorov2016}.

In this work, we present the first implementation of a circuit-based superconducting sensor for the detection of the mechanical motion of a levitated superconducting sphere \cite{RomeroIsart2012, Gutierrez2022, Gutierrez2023, Hofer2023}. The mechanical system provides extremely long dephasing times combined with a comparatively large levitated mass, thereby offering the prospect of engineering  quantum superpositions in the high-mass regime \cite{Cirio2012, RomeroIsart2012, Reed2017, Pino2018, Bild2023, Hofer2023,Schrinski2019}. Such systems are promising candidates for delocalized quantum source masses in gravity experiments, and for acceleration sensors in search for new physics beyond the standard model \cite{Bose2017, RomeroIsart2011, DeWitt2011, Wagner2012, Marletto2017, Burrage2018, Marshman2020, Brax2022, Aspelmeyer2022, Moore2021, Carney2021, Higgins2024, Richman2023}.

For the detection of the center of mass motion, we integrate a SQUID in a microwave resonator and couple it inductively via a flux transformer to a pick-up loop (PUL), which is placed in the vicinity of a levitated superconducting microsphere. The motional coupling is sufficiently strong to read out the mechanical motion with single-digit microwave photon numbers, a regime which will make it directly utilizable with quantum logic circuits for non-classical state generation \cite{Hofheinz2008, OConnell2010, Cirio2012, Pirkkalainen2015, Streltsov2021}.

\section{Experiment}
\begin{figure*}
  \includegraphics[scale=1]{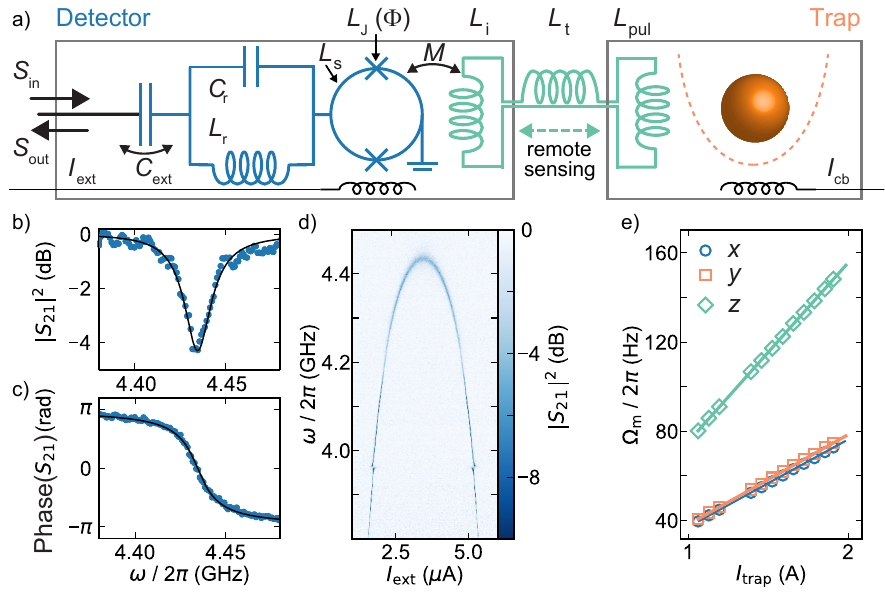}
        \caption{Panel a) The flux-tunable cavity (blue) is coupled to the motion of a levitated particle (orange) via a flux transformer circuit (green). The flux transformer contains a pickup loop, which is exposed to the motion of the superconducting particle (brown) trapped in a magnetic field gradient (orange). The individual parts are characterized in the lower panels: b) and c) display amplitude  and phase of the complex microwave scattering parameter of the flux-tunable cavity. Panel d) depicts the tunability over a range of over $\SI{600}{\mega\hertz}$ before it drops below our HEMT working range. At the limits of this range, tunabilities of up to $\SI{4}{\giga\hertz}/\Phi_0$ were achieved. In panel e) we determine that the translational modes of the levitating superconductor exhibit a linear growth with the trap current. The maximum observed frequency is $\Omegam / 2\pi = \SI{160}{\hertz}$ in the vertical direction $z$. In the trap's planar $xy$ plane, the field gradient is half as high, resulting in reduced frequencies. Error bars are smaller than the marker sizes.}
        \label{fig:Schematics}
\end{figure*}
 \subsection{Overview}\label{subsec:overview}
The experiment is schematically represented in Fig.~\ref{fig:Schematics}a: 
The detection is realized by a flux-tunable microwave cavity with eigenfrequency $\wres(\Phi)$, that consists of a coplanar waveguide resonator shunted to ground via a SQUID loop of two Josephson inductances $\Lj$ and geometric loop inductance $\Ls$ ($\ll \Lj$). External magnetic flux $\Phi$ tunes the Josephson inductance with a period given by the flux quantum, $\Phi_0 = \SI{2.068e-15}{\tesla\square\metre}$ \cite{Clarke2004}. 
Spectroscopically, this mechanism is visible in the bias field dependence of the reflection properties of the microwave resonator. An external coil is used to apply a magnetic flux bias. At a fixed bias, the resonance is visible in the scattering parameter $S_{21} = S_{\mathrm{out}}/S_{\mathrm{in}}$ of the vector network analyzer (VNA), as depicted in Fig.~\ref{fig:Schematics}b and c. The dependence of the resonance frequency on the bias is shown in Fig.~\ref{fig:Schematics}d. From this measurement, we extract a flux response $\slope \equiv \partial\wres/( 2\pi \partial \Phi)$ of up to $ \SI{4}{\giga\hertz}/\Phi_0$ (see also Fig. \ref{fig:FTR_details} in the Appendix).

A flux transformer forms the link between the SQUID and the PUL near the levitated microsphere (highlighted in green in Fig.~\ref{fig:Schematics}a). The magnetic field lines of the trap bend around the microsphere, and so as the sphere moves, it changes the magnetic flux threading the PUL. The transformer contains an input coil with inductance $\Lfc$, mounted on top of the SQUID loop. The flux transfer between the input coil and the SQUID loop is determined by their mutual inductance $\Mi = \nui\sqrt{\Ls\Lfc}$, including the input coupling coefficient $\nui$. A twisted pair wire with an inductance $\Ltw$ connects the input- and PUL-coils, where the latter has an inductance of $\LPUL$. The parameter $\atech = \Mi/(\Lfc + \Ltw + \LPUL) \approx 10^{-3}$ quantifies how efficiently the flux is coupled from the PUL into the SQUID ring (see  Appendix \ref{app:pickup_position}).

The field disturbance caused by the particle depends on its radius $\rp$ and the magnetic field gradient $b_{i}$ along the translational degrees of freedom $i \in {x, y, z}$ of the particle. The disturbance induces a flux in the PUL $\partial \Phim / \partial i =  \Ff_{i} b_{i} \rp^2$, where we introduce a geometric coupling factor $\Ff_{i}$. This factor strongly depends on the separation between the pickup and the trap centre for each translational mode (here $\approx(4.5, 2.5, 2.5)\cdot10^{-4}\,\mathrm{m}$ in $x$, $y$, and $z$ direction, for details see Appendix \ref{app:pickup_position}).
For the trap parameters applied in the experiments shown in Fig.~\ref{fig:CouplingScaling}, we measure $\Ff = (5.7, 67, 6.1)\cdot10^{-4}$ for the directions $x$, $y$, and $z$.

The trap configuration used for magnetic levitation of the superconducting microsphere is described in detail in Ref. \cite{Hofer2023}. It consists of two slightly elliptical superconducting coils in an anti-Helmholtz like configuration with a current of magnitude $\Itrap$, which creates field gradients $b_{i}$. The sphere's oscillation frequencies are given by
\begin{equation}
\Omegam^i = \sqrt{\frac{3}{2\muN\rho}} |b_{i}|,
\label{eq:trapfreq}
\end{equation}
where $\rho = \SI{1.1e4}{\kilo\gram\per\metre^3}$ is the density of the lead-tin particle and $\muN$ is the vacuum permeability. The trap is mechanically suspended for vibration decoupling and is magnetically shielded by a superconducting aluminum box. The particle has a radius $\rp = \SI{50}{\micro\metre}$ and mass $\mtot = \SI{5.7}{\micro\gram}$. For typical trapping frequencies around $\SI{100}{\hertz}$, the amplitude of the quantum mechanical zero-point motion of the harmonic oscillator $\xzpf^i = \sqrt{\hbar / (2\mtot\Omegam^i)}$ is about $\SI{4}{\femto\metre}$. The particle displacement $\zeta_i(t)$ and translational frequency $\Omegam^i$ are recorded using a camera to calibrate the detector response of the flux tunable resonator. The linear dependence of frequency $\Omegam^i$ on trap current is shown in Fig. \ref{fig:Schematics}e, yielding a magnetic field gradient of $23.5, 24.2$ and $\SI{48.1}{\tesla \per \metre}$ per applied ampere of $\Itrap$ in the $x$, $y$, and $z$-directions, respectively. A second, smaller coil is placed close to the particle to apply oscillating magnetic fields which is also used for calibration. More details on the setup are provided in Appendix \ref{sec:DetailedSetup}, including fabrication of the microchips used in the experiment.\\

The inductive electromechanical coupling is described by the interaction Hamiltonian $\Hint = -\hbar \gNull \agen \akill(\bgen + \bkill)$, with the ladder operators $\akill$ $(\bkill)$ of the microwave resonator (mechanical oscillator)  \cite{Rodrigues2019, Schmidt2020, Zoepfl2020, Bera2021}. The electromechanical single-photon coupling strength is given by \cite{RomeroIsart2012, Hofer2023}
\begin{equation}
\gNull^i \equiv \Gmech_i \xzpf = \frac{\partial \wres}{\partial \Phi}\frac{\partial \Phi}{\partial i}\xzpf^i = 2\pi\slope \atech\Ff_i \bi  \rp^2 \xzpf^i ,
\label{eq:gNull}
\end{equation}
where we have defined the electromechanical coupling strength $G_i=\partial \wres / \partial i$, described in detail below. For our parameters the interaction strength is between $0.1-\SI{10}{\milli\hertz}$, as it is different for each mechanical degree of freedom. We emphasize that the overall scaling of the coupling strength versus particle radius is $\rp^{1/2}$, since the radius also affects $\xzpf \propto \rp^{-3/2}$ via the particle mass, rendering this technique suitable for quantum state generation of heavy objects. In addition, the coupling can be controlled {\it in situ} by tuning the magnetic field gradient $\bi$ of the trap or the bias point of the cavity, which affects $\partial \wres / \partial \Phi$. 

\subsection{Electromechanical coupling}\label{subsec:spectrum}
\begin{figure}
  \includegraphics[scale=0.97]{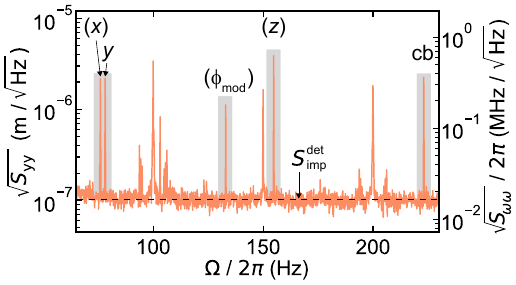}
        \caption{The sphere's displacement spectral density and relevant peaks highlighted by a grey background, calibrated for the $y$-direction. For this motion an imprecision level of $\Simp = \SI{102}{\nano\metre\per\sqrt{\hertz}}$ is found. The modes corresponding to the $x$ and $z$ directions are also detected, but require individual displacement calibration. A modulation tone applied to the calibration coil (cb) at \SI{223}{\hertz} allows us to calibrate the frequency spectral density (right axis). A further applied phase modulation tone, labelled $\phi_{\mathrm{mod}}$, is visible at $\SI{133}{\hertz}$. The read-out strength corresponds to an average photon number of $\nr = 20$. Noise peaks from the electric grid (multiples of $\SI{50}{\hertz}$) and upconverted low-frequency noise around these features are also visible (not highlighted).}
        \label{fig:DisplacementExample}
\end{figure}
The motion of the levitated particle is monitored via the microwave cavity by applying a microwave probe tone resonant to the cavity at frequency $\wres$. The reflected signal is down-converted, digitized, and transformed into frequency space. To calibrate the signal trace, the motional amplitude of the particle is determined by optically recording its motion via a camera, resulting in the displacement spectral density $S_{ii}$ \cite{Hofer2023}. In addition, a calibrated flux modulation tone is applied via the calibration coil, to obtain the flux noise spectral density $\Sphiphi$. Using the flux response of the cavity one can derive the frequency noise spectral density as $\Sww = \Sphiphi  (2\pi \slope)^2$. 

An example of a calibrated spectrum is shown in Figure \ref{fig:DisplacementExample}. Here, the particle amplitude in the $y$-direction was $\SI{740}{\nano\metre}$, far above the thermally driven amplitude ($\approx \SI{10}{\pico\metre}$). This excitation is induced during the transport from the initial resting position at a distance of over $\SI{2}{\milli\metre}$ from the final trap centre.
From the displacement spectrum we find a detected off-resonance displacement noise spectral density (imprecision level) of $(\SimpD)^{1/2} = \SI{102}{\nano\metre\per\sqrt{\hertz}}$ (black dashed line). We underline that the calibration is different for each oscillation mode of the particle, bias flux setting, and probing power. The $x$ and $z$ modes, labeled in brackets, thus require individual calibration. \\
\begin{figure}
  \includegraphics[scale=1]{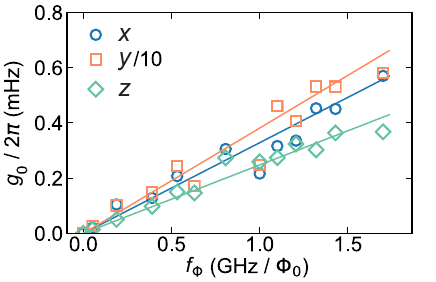}
        \caption{The microwave cavity enables {\it in situ} tuning of the electromechanical coupling by adjusting its bias field, thus affecting $\slope$. We observe a direct proportionality as predicted by Eq.~(\ref{eq:gNull}). The $y$-mode coupling is rescaled by a factor of ten, since it couples about ten times more strongly than the other modes, which is caused by the relative position of the PUL to the sphere (see text).}
        \label{fig:CouplingScaling}
\end{figure}
The flux calibration tone at $\SI{223}{Hz}$ is applied as a current $\Ifb$ via the calibration coil. It induces a calibrated amount of flux, from which we extract $\Sphiphi$. By using the flux response of the tunable resonator, here biased at $\slope = \SI{188}{\mega\hertz}/\Phi_0$, we determine the frequency noise spectral density $\Sww$ (right axis). Finally, the electromechanical coupling rate for this setting is
\begin{equation}
    \frac{\Gmechy}{2\pi} = \sqrt{\frac{\Sww (\Omegam)}{2\pi\Syy (\Omegam)}} = \frac{\SI{0.35}{\mega\hertz\per\sqrt{\hertz}}}{\SI{2.2}{\micro\metre\per\sqrt{\hertz}}} = \SI{0.16}{\tera\hertz\per \metre},
\end{equation}
corresponding to a vacuum coupling strength of $\gNull^{\mathrm{y}} / 2\pi  = \SI{0.7}{\milli\hertz}$ in this particular measurement. See App.~\ref{sec:calibofcalibcoil} for details of the calibration procedure. 

Our setup allows to tune the electromechanical coupling via $\slope$ by  varying the bias of the microwave cavity, cf. Eq.~(\ref{eq:gNull}).
This bias can be controlled by applying a current to the external bias coil in the chip assembly (see Fig.~\ref{fig:Schematics}), enabling fast control of the coupling strength (i.e. faster than $\Omegam$), while keeping the trapping conditions unaffected. The observed relationship between the flux response and the coupling strength, shown in Fig.~\ref{fig:CouplingScaling} for static $\slope$, confirms the expected linear dependence. For a trap gradient of $\bi = (39, 40, 80)\,\mathrm{T}/\mathrm{m}$, corresponding to $\Omegam^i / 2\pi = (69,70, 140)\,\mathrm{Hz}$ and $\xzpf^i = (4.6, 4.6, 3.2)\,\mathrm{fm}$, we determine slopes corresponding to $\gNull^i / (2\pi \slope) = (0.33, 3.8, 0.25) \,\mathrm{mHz} / (\mathrm{GHz}/\Phi_0)$, within the predicted range for the present implementation. The electromechanical coupling corresponds to an induced flux of $\partial \Phiind / \partial i = (70, 800, 80)\,\Phi_0/\mathrm{m}$.

The flux coupling (summarized by the parameter $\Ff_i$) depends highly on the relative position of the sphere with respect to the PUL. The ratios of the coupling strengths to each other are found to be consistent with distances of $(450, 250, 250)\,\mu \mathrm{m}$, using analytical calculations of the flux coupling. Using the optical access, the $z$-spacing was found to be $\SI[separate-uncertainty=true]{300(50)}{\micro\metre}$, which provides an independent confirmation of our estimates. The $x$ and $y$ displacements could not be determined from the recorded images. In order to match the absolute coupling strength, we then require $\atech = 5\cdot 10^{-3}$, which is of the same order as the estimated transfer inductance and the alignment of the input coil. Hence, the experimental coupling is broadly consistent with the expected value, though we underline that the interdependent calculation of $\atech$ and $\Ff$ are both subject to some uncertainty. Details on on the modelling of the flux coupling can be found in Appendix \ref{app:pickup_position} and Refs. \cite{Hofer2019, Hofer2023}.\\

\subsection{Imprecision level \& detection efficiency}
\label{subsec:imprecision}
\begin{figure}
  \includegraphics[scale=0.8]{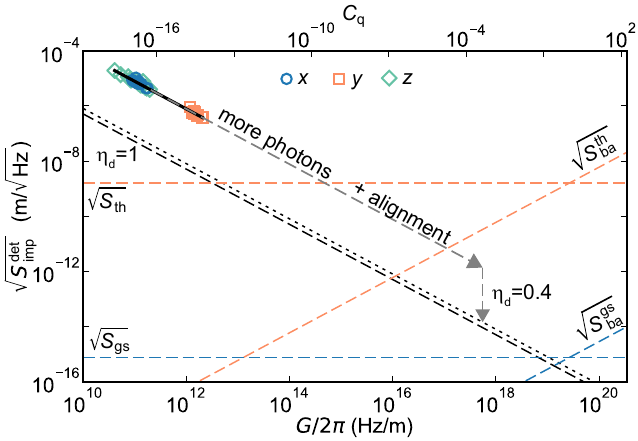}
        \caption{On-resonance ($\Omega = \Omegam$) electromechanical readout performance. The experimental imprecision values for the $x$, $y$, and $z$ modes are extracted from coupling measurements (see also Appendix \ref{sec:optomech_app} for a detailed zoom in the data regime). A detection efficiency of $4.3\cdot10^{-5}$ is extracted from the enhanced imprecision level (black solid line). Further, the thermal displacement ($S_{\mathrm{th}}$, dashed) and back-action noise density in thermal equilibrium with the cryogenic environment ($S_{\mathrm{ba}}^{\mathrm{th}}$, dashed) are displayed in orange. Upon cooling to the ground state, electromechanical damping is induced, broadening and thus reducing the on-resonance back-action level ($S_{\mathrm{ba}}^{\mathrm{gs}}$, dashed, blue). To achieve this cooling the measurement needs to resolve the displacement of a single excitation $S_{\mathrm{gs}}$ for the particular mode, here along the strong trap axis $z$ (blue, dashed). Dashed arrows indicate the improvement feasible with better alignment of the PUL to the sphere, and the improvement given by a quantum limited amplifier to reach $\etad = 0.4$. Further improvements are detailed in the text. Error bars are smaller than the marker size.}
        \label{fig:CavityOptomech}
\end{figure}
A key parameter of the electromechanical detector is its imprecision level. It can be described using the formalism developed for cavity optomechanical systems where the optical (mechanical) subsystem's loss rate is denoted by $\kappa$ ($\Gammam$) \cite{Aspelmeyer2014}. There, the imprecision level is given by $\SimpC = \kappa/(16 \nr G^2)$ where we assume the absence of photon loss, added noise from our detection chain, and that the system is far from the sideband-resolved regime ($\Omegam \ll \kappa$).  The effective coupling rate ($\sqrt{\nr} \gNull$) and the system loss rates $\kappa$ and $\Gammam$ give the quantum cooperativity $\Cq = 4\nr \gNull^2/(\kappa\Gammam\nth)$, where $\nth\approx  \kB T / (\hbar \Omegam)$ is the thermal phonon occupation and $\kB$ is the Boltzmann constant. The imprecision level can then be written in terms of the cooperativity,  $\SimpC = \Sgs / \Cq$, where $\Sgs = 4 \xzpf^2 / \Gammam \nth$ corresponds to resolving displacements on the scale of the mechanical ground state extent on a timescale comparable to the thermal decoherence time. The quantum cooperativity also allows to define a measurement efficiency, $\etae =(1+1/\Cq)^{-1}$. The observed imprecision level $\SimpD$ will be equal to or greater than $\SimpC$ due to limitations of the apparatus. This increase is quantified by the detection efficiency $\etad = \SimpC / \SimpD$. Together, these two efficiencies yield the total measurement efficiency, $\eta = \etad \etae$ \cite{Clerk2003, Magrini2021}.

The best value for  $\SimpD$ obtained in this first-generation device is $\SI{100}{\nano\metre\per\sqrt{\hertz}}$, with a readout photon number around $\nr \approx 20$. To investigate the cavity's noise floor, we apply a weaker probe power to reach $\nr = 0.05 \pm 0.03$, thus deep in its linear coupling regime \cite{Pirkkalainen2015} and we set the flux sensitivity at $\slope = \SI{1.7}{\giga \hertz}/\Phi_0 $ via $\Iext$. We then tune the interaction strength by varying the trap gradient $\bz$, and determine the coupling strength $G$ and imprecision level for each value of $\bz$. The results are plotted for all three mechanical modes in Fig.~\ref{fig:CavityOptomech}. The coupling strength differs for the individual modes as they couple differently to the PUL, cf. Appendix \ref{sec:Elmechconversion}. Nonetheless, we find that the coupling and imprecision of all modes follow $\SimpD \cdot (G/2\pi)^2 = \SI[separate-uncertainty = true]{0.61(2)}{\tera\hertz} $ (black solid line; see Appendix \ref{sec:optomech_app} for a magnified view of the data range). The quantum-limited noise density for perfect detection is $\SimpC \cdot (G/2\pi)^2 = \SI[separate-uncertainty=true]{26(13)}{\mega\hertz}$. Thus we find $\etad = (4.3\pm2.1)\cdot10^{-5}$. We attribute the decreased efficiency to signal lost from the cavity, added noise at the high-electron mobility transistor (HEMT) amplifier, and imperfect room temperature amplification (see Appendix \ref{sec:detefficiency} for details).

\section{Outlook}\label{sec:outlook}
\subsection{Path towards the quantum regime}\label{subsec:quantum}
Following the description in the previous section, the detection efficiency determines the imprecision level that can be reached. In turn, this resolution gives the phonon number that the mechanical system can be directly cooled to, with sufficiently strong feedback, and is given by $\nmmin = (1/\sqrt{\eta} - 1)/2$ \cite{Magrini2021}. Therefore $\eta > 1/9$ is required for ground state cooling. For our system, this value corresponds to a noise spectral density below $\Sgs = (\SI{0.8}{\femto\metre})^2/\mathrm{Hz}$ highlighted in Fig.~\ref{fig:CavityOptomech} as a blue dashed line for the $z$-mode \cite{Magrini2021}. Thus, it will be necessary to increase the coupling strength and reduce the detection noise. To quantify the necessary improvements, we focus only on the $z$-mode along the strong trap axis, as it enables higher frequencies and coupling rates. In addition, we assume to reach similar mechanical properties as discussed in an earlier work employing an identical trap design. Namely $T = \SI{15}{\milli\kelvin}$ and $Q = \Omegam / \Gammam = 2.6\cdot10^7$ \cite{Hofer2023}.

We also conservatively consider all loss channels contributing to the thermal decoherence, thus upon ground state cooling the effective linewidth broadens to $\Gammam \rightarrow \Gammaeff = \Gammam \nth$. Describing the displacement noise spectral density of the detector requires the back-action force noise spectral density $S^{\mathrm{ba}}_{\mathrm{FF}} = 4\hbar^2G^2\nr/ \kappa$ (in the limit $\Omegam \ll \kappa$) to be multiplied by the mechanical susceptibility $\chim (\Gamma) = (m (\Omegam^2 - \Omega^2 - i\cdot \Gamma \Omega))^{-1}$. This requires to distinguish between the back-action force noise of an element in equilibrium with the thermal environment, having a linewidth of $\Gammam$ and an element cooled to its quantum mechanical ground state with broadened linewidth $\Gammaeff$. In particular, we define the on-resonance ($\Omega \rightarrow \Omegam$) back-action displacement noise spectral densities $S_{\mathrm{ba}}^{\mathrm{th}} = S^{\mathrm{ba}}_{\mathrm{FF}}|\chim(\Gammam)|^2$ and  $S_{\mathrm{ba}}^{\mathrm{gs}} = S^{\mathrm{ba}}_{\mathrm{FF}}|\chim(\Gammaeff)|^2$.\\
The microwave sensing implementation enables the use of squeezed light to enhance the detection efficiency, in particular by using Josephson parametric amplifiers. For quantum-limited amplifier systems, $\etad$ up to $0.7$ has been demonstrated \cite{Renger2021}. Implementing such a device with a critically coupled microwave cavity would allow us to reach efficiencies of up to $\etad = 0.4$ for $\SI{30}{\dB}$ gain, achievable using a non-degenerate parametric amplifier (black dotted line in Fig.~\ref{fig:CavityOptomech}) \cite{Winkel2020}, cf. Appendix \ref{sec:detefficiency}. With such a detection efficiency, the electromechanical quantum cooperativity $\Cq$ required to reach $\eta > 1/9$ is $0.38$.\\ 
In the measurement presented in Fig.~\ref{fig:CavityOptomech} we achieve a quantum cooperativity of $\Cq = 5\cdot10^{-17}$ at a readout level of $\nr = 0.05$. While this presents a significant gap towards the parameters required for quantum control, we are confident that this can be achieved. The experimental factors which contribute to the small cooperativity in this first-generation device offer multiple opportunities for improvement. In particular, the flux coupling can be significantly increased: Firstly, {\it in situ} tuning of the position of the sphere relative to the PUL, affecting $\Ff$, provides a predicted enhancement in cooperativity by a factor $4\cdot10^{10}$. This fine-tuning can be achieved through the utilization of nano-positioners, which can adjust the position of the PUL to bring it closer to the particle. Alternatively, the particle can be shifted closer to the PUL by manipulating the trap center with auxiliary coils. This improvement, together with the use of a higher readout photon number $\nr = 10$, are highlighted by the downward diagonal arrow in Fig.~\ref{fig:CavityOptomech}. Secondly, $\atech$ can be improved by reducing the parasitic inductance $\Ltw$ from the flux transformer and by better alignment between the input coil and the SQUID loop. We find that the enhanced flux transformation will translate to an increase of $\Cq$ by a factor of $2.5\cdot10^{3}$. Together, these improvements will be sufficient to enable ground-state cooling. We list further possibilities for improvement of about four orders of magnitude, relating to the cavity parameters, in Appendix \ref{sec:optomech_app}. In addition, different levitator geometries, e.g. microrings, can be beneficial to enhance the coupled flux \cite{Navau2021}. The presented coupling can also be used to couple the levitated object to superconducting qubit circuits\cite{Ilichev_2007,Bal2012,Toida2023}, enabling the implementation of quantum sensing methods for nonclassical state preparation \cite{RomeroIsart2012, Streltsov2021}.

\subsection{Conclusions}\label{subsec:conclusions}
In summary, we have implemented and characterized a cavity-based scheme to read out the motion of a magnetically levitated, superconducting microsphere using a superconducting, flux-tunable microwave resonator. Our experiment successfully demonstrates {\it in situ} control over the electromechanical coupling strength via the trap strength and the cavity bias field. The imprecision noise in our setup is presently limited by signal loss to the cryogenic microwave amplifier, which has a noise temperature of $\SI{2.5}{\kelvin}$, and by imperfections in the room temperature electronics used at present.
These findings allow to describe a clear path towards the quantum regime of mechanical motion for microgram-scale masses. 

The data of this study are available at the Zenodo repository \cite{data}.

\paragraph{\textbf{Acknowledgements}}
We gratefully acknowledge valuable discussions with Uros Delic, Lorenzo Magrini, Corentin Gut. This work was supported by the European Union’s Horizon 2020 research and innovation programme under Grant No. 863132 (iQLev) and 101080143 (SuperMeQ), the European Research Council under Grant No. 951234 (ERC Synergy QXtreme), by the Austrian and Bavarian Academy of Sciences (Topical Team SGQ), by the Alexander von Humboldt Foundation through a Feodor Lynen Fellowship (P.S.), the Swedish Research Council under Grant 2020-00381 (G.H.), and  the Deutsche Forschungsgemeinschaft (DFG, German Research Foundation) via Germany’s Excellence Strategy EXC-2111- 390814868 (H.H., R.G.).

\appendix
\label{sec:_SampleParameters}
\begin{table*}
\begin{center}
    \begin{tabular}{rlrc}
    \hline
 &  Parameter &  Value  & Comments \\
    \hline
  \multirow{13}{*}{\tabrotate{\textbf{CPW-FTR}}}	&	Sweet spot frequency & $ \wres /2\pi=   \SI{4.44}{\giga\hertz}$ & \\
														&	Sweet spot intrinsic linewidth & $ \kint /2\pi=   \SI{5}{\mega\hertz}$ & no remote sensor connected \\
														&	External microwave coupling & $ \kext /2\pi=   \SI{18}{\mega\hertz}$ & \\
                                                     	  & Bare cpw eigenfrequency & $\wrN/2\pi =  \SI{7.77}{\giga\hertz}$ & \\
														&	CPW length & $\lr =  \SI{3.8}{\milli\metre}$ & \\
														&	CPW dimensions (width, gap) & $ \widthr= \SI{10}{\micro\metre}$ \ \ \  $\sr= \SI{6}{\micro \metre}$ & \\
														&	CPW resonator inductance, capacitance & $ \Lr =   \SI{1.4}{\nano\henry}$ \ \ \ $\Cr =\SI{310}{\femto\farad}$ &  lumped element equivalent\\
														&	CPW impedance & $ Z =   52\,\Omega$ & \\
														&	Effective dielectric constant & $ \eeff = 6.45$ & \\
    \hline
      \multirow{9}{*}{\tabrotate{\textbf{SQUID}}}	 	  &	SQUID inductance & $ \Ls =  \SI{0.12}{\nano\henry}$  & simulated \\
														&	Josephson inductance & $ \Lj =  \SI{0.36}{\nano\henry}$ & at sweet spot \\
														&	Critical current & $ \Ics \approx \SI{0.5}{\micro\ampere}\,$ & single junction \\
														&	Washer outer diameter & $ \Ws = \SI{125}{\micro\metre}$ &  \\
              											&	Washer inner diameter & $ \Ds = \SI{25}{\micro\metre}$ &  \\
                         								&	Effective area & $ \Aeff = $(\SI{56}{\micro\metre})$^2$ & derived from \cite{Gross1990} \\
														&	Screening parameter & $ \betaL =  0.06 \,$ &  \\
	\hline
       \multirow{13}{*}{\tabrotate{\textbf{Flux trans.}}}	 	  &	Input coil outer length, Windings & $ \Wi = \SI{210}{\micro\metre} $ \ \ \ $\Numberi = 14$ &  \\
														& Input coil conductor (width, gap) & $ \widthi =  \SI{3.5}{\micro\metre}$ \ \ \  $\Si =  \SI{3.5}{\micro\metre}$ &  \\
														& Input coil inductance & $ \Lfc =  \SI{20.5}{\nano\henry}\,$ & Wheeler's equation \\
														&	Squid-input coil spacing vertical, radial & $ \Delta z_i \approx \SI{30}{\micro\metre}$ \ \ \ $\Delta r_i \approx \SI{60}{\micro\metre}$ & flip chip geometry  \\
              											&	Input coupling coefficient, mutual inductance & $ \nui \approx 0.1$ \ \ \ $\Mi = \SI{0.16}{\nano\henry}$ & simulated using $\Delta z_i$, $\Delta r_i$  \\
                                                  		&	PUL single loop diameter, loop centers separation & $ \DPUL = \SI{150}{\micro\metre}$ $\SPUL = \SI{158}{\micro\metre}$ &  \\
                         								&	PUL loop inductance & $ \LPUL = \SI{0.9}{\nano\henry}$ & simulated \\
														&	Flux transportation inductance & $ \Ltw \approx \SI{100}{\nano\henry}$ &  estimation based on Ref.~\cite{Hofer2023}\\
              											&	Flux transfer efficiency & $ \atech = 1.3 \cdot10^{-3} (5 \cdot10^{-3})$ &  design (fitted from data) \\
	\hline
\multirow{11}{*}{\tabrotate{\textbf{Trap}}} & Mechanical eigenfrequency & $\Omegam/2\pi =  (39, 40, 80)\,\mathrm{Hz} / \Itrap$ & cf. Fig.~\ref{fig:Schematics}d) \\
                                                        & Trap gradient per applied current & $\BI = (23.5, 24.2, 48.1) \mathrm{T m}^{-1} / \Itrap$ & measured in Fig.\,\ref{fig:Schematics}e\\
														& Maximum trap current & $\Itrap \approx \SI{8}{\ampere} $ & limit by critical current density\\
														& Mechanical quality factor & $Q =  2.6\cdot10^7$ & measured in Ref. \cite{Hofer2023}\\
														& Sphere radius & $ \rp = \SI{50}{\micro\metre} $ & \\
              											& Sphere density & $ \rhop = \SI{10.9e3}{\kilogram\per\cubic\metre} $ & 90:10 lead-tin alloy\\
														& Sphere mass & $ \mtot = \SI{5.7e-9}{\kilogram} $ & \\
														& Zero-point motion & $ \xzpf = \SI{4}{\femto\metre}$ & typical (frequency dependent)\\
              											& Flux input coupling & $ \Ff = (5.7, 67, 6.1)\cdot10^{-4}$ & for trap conditions in Fig.~\ref{fig:CouplingScaling}\\
                                                        & Total flux transformation & $\partial \Phiind / \partial i = (70, 800, 80) \Phi_0 / \mathrm{m} $ & for trap conditions in Fig.~\ref{fig:CouplingScaling} \\
                                       					& Thermal phonon occupation limit & $ \nm^{\mathrm{th}} = 2\cdot10^{6} $ & at $\SI{150}{\hertz}$ and $\SI{15}{\milli\kelvin}$\\
	\hline
    \end{tabular}
\caption{\textit{Overview of the experimental device parameters.}}
\label{tab:SampleParameters}
\end{center}
\end{table*}
\section{Experimental details}
\label{sec:DetailedSetup}
\subsection{Coupling details}
The experiment is depicted in Fig.~\ref{fig:FullSetup} and \ref{fig:Schematics} of the main text. A $\mtot \ \approx \ \SI{6}{\micro\gram}$ lead-tin sphere of radius $\rp = \SI{50}{\micro\metre}$ is levitated in a magnetic field gradient $b_i$ with $i \in x,y,z$ the coordinates of the trap. For simplicity, we focus on the vertical axis, parallel to the gravity axis, hence \ps{$i = z \rightarrow \ $}$\bz$, $\Omegam^z$, etc. The other modes can be described analogously. The sphere acts as a linear oscillator with frequency $\Omegam^z$ and displacement $ \zop = \xzpf^z (\bgen + \bkill)$, described semi-classically by the phonon creation operator $\bgen$, phonon annihilation operator $\bkill$, and the zero-point motion $\xzpf^z = \sqrt{\hbar/(2\mtot\Omegam^z)}$. The motion of the sphere $\delta z$ along the $z$-axis changes the flux within the superconducting PUL $\delta \Phim^z = \Ff\bz\rp^2 \delta z$. Here, $\Ff$ describes how effectively the motion is coupled into the loop. As the flux within the transformer remains, screening currents are generated to counter $\delta\Phim$. The currents' strengths depend on the inductances of the loop, consisting of the detecting PUL inductance $\LPUL$, the accumulated inductances to guide the current to the detector $\Ltw$, and an input coil at the detector $\Lfc$, mounted on top of it in a flip-chip configuration. A fraction of these induce a flux $\delta \Phiind^z $ in the SQUID loop of the microwave cavity. The loop efficiency $\Phiind / \Phim$ is denoted by $\atech = \Mi / (\LPUL + \Lfc + \Ltw)$, independent of the motion axis. Here, $\Mi$ describes the mutual inductance between flip chip coil and SQUID loop $\Mi = \nui \sqrt{\Lfc \Ls}$ with the input coupling coefficient $\nui$. Thus, a mechanical motion of the sphere with amplitude $\delta z$ induces a flux in the SQUID loop of the electric resonator as $\atech\Ff\bz\rp^2 \delta z$. Experimentally, the oscillating sphere will give rise to an rms flux noise peak which can be related to its phonon number $\nm = \langle \bgen\bkill \rangle$:
\begin{equation}
\langle \Phiind^z \rangle ^{\frac{1}{2}} = \atech\Ff\bz\rp^2 \xzpf^z \sqrt{(2\nm + 1)}.
\label{eq:meanFlux}
\end{equation}
The flux induced in the SQUID will shift the resonance frequency of the flux-tunable cavity by 
\begin{equation}
\delta \omega = \frac{\partial \wres}{\partial \Phi} \delta \Phiind^z.
\label{eq:freqshift}
\end{equation}
In leading order this gives rise to an electromechanical coupling as $\wres(\zop) = \left( \wres - \Gmech \zmean \right)$, where $\Gmech$ = $\partial \wres / \partial z$. Thus the interaction Hamiltonian becomes:
\begin{equation}
\Hint = -\hbar \gNull \agen \akill \left(\bgen + \bkill  \right).
\label{eq:Hint}
\end{equation}
The vacuum coupling strength $\gNull = G\xzpf$ is derived to be:
\begin{equation}
\gNull = \frac{\partial \wres}{\partial \Phi}\frac{\partial \Phiind}{\partial z}\xzpf = \frac{\partial \wres}{\partial \Phi} \atech\Ff\bz\rp^2 \xzpf.
\label{eq:gNulllong}
\end{equation}

\subsection{Microchip fabrication}
\label{sec:Microchips}

\paragraph{Flux-tunable microwave cavity}
The cavity fabrication begins by using a commercial high resistivity silicon wafer ($> \SI{10}{\kilo\ohm\centi\metre}$), which is diced into 20$\times\SI{20}{\milli\metre}$ batches. After wet cleaning, a double layer resist is spin coated on top. The whole electronic structure, consisting of coupling capacitance, coplanar waveguide and Josephson junctions, is then patterned using electron beam lithography. Then the batch is developed and two aluminum layers of thickness $\SI{30}{\nano\metre}$ and $\SI{60}{\nano\metre}$ are grown using shadow-angle electron beam evaporation. Before the evaporation, a mild ion etching is applied and in between the evaporations, the first layer is oxidized for $\SI{5}{\minute}$ at an oxygen pressure of $\SI{5}{\milli\bar}$, generating the Josephson tunneling barrier. Then a lift-off process is applied and the batch is finally diced into 5$\times\SI{2}{\milli\metre}$-sized individual cavity chips.

\paragraph{Gradiometric pick-up loop}
We use a high resistivity silicon two inch waver and sputter $\SI{300}{\nano\metre}$ of niobium on top. Using optical lithography with a mask-less aligner and SF$_6$ etching, the gradiometric pick-up loops are patterned into the niobium. Finally, the structures are diced into 45$\times\SI{5}{\milli\metre}$ individual chips.

\paragraph{Flip chip coil}
For the flip chip coil, the first step is the same as for the pick-up loop except the thickness of the niobium is $\SI{75}{\nano\metre}$. Using optical lithography with a mask-less aligner and SF$_6$ etching, the multiturn coil and the supply lines with one disconnection to the middle of the coil are patterned into the niobium film. We use Plasma Enhanced Chemical Vapor Deposition (PECVD) to coat $\SI{150}{\nano\metre}$ of Si$_3$N$_4$ as dielectric and a second niobium layer of $\SI{180}{\nano\metre}$ to connect the middle of the coil to the supply line. For both steps lift off is used for structuring.

\subsection{Cryogenic setup assembly}
\label{sec:Assembling}

\begin{figure}
  \includegraphics[scale=0.5]{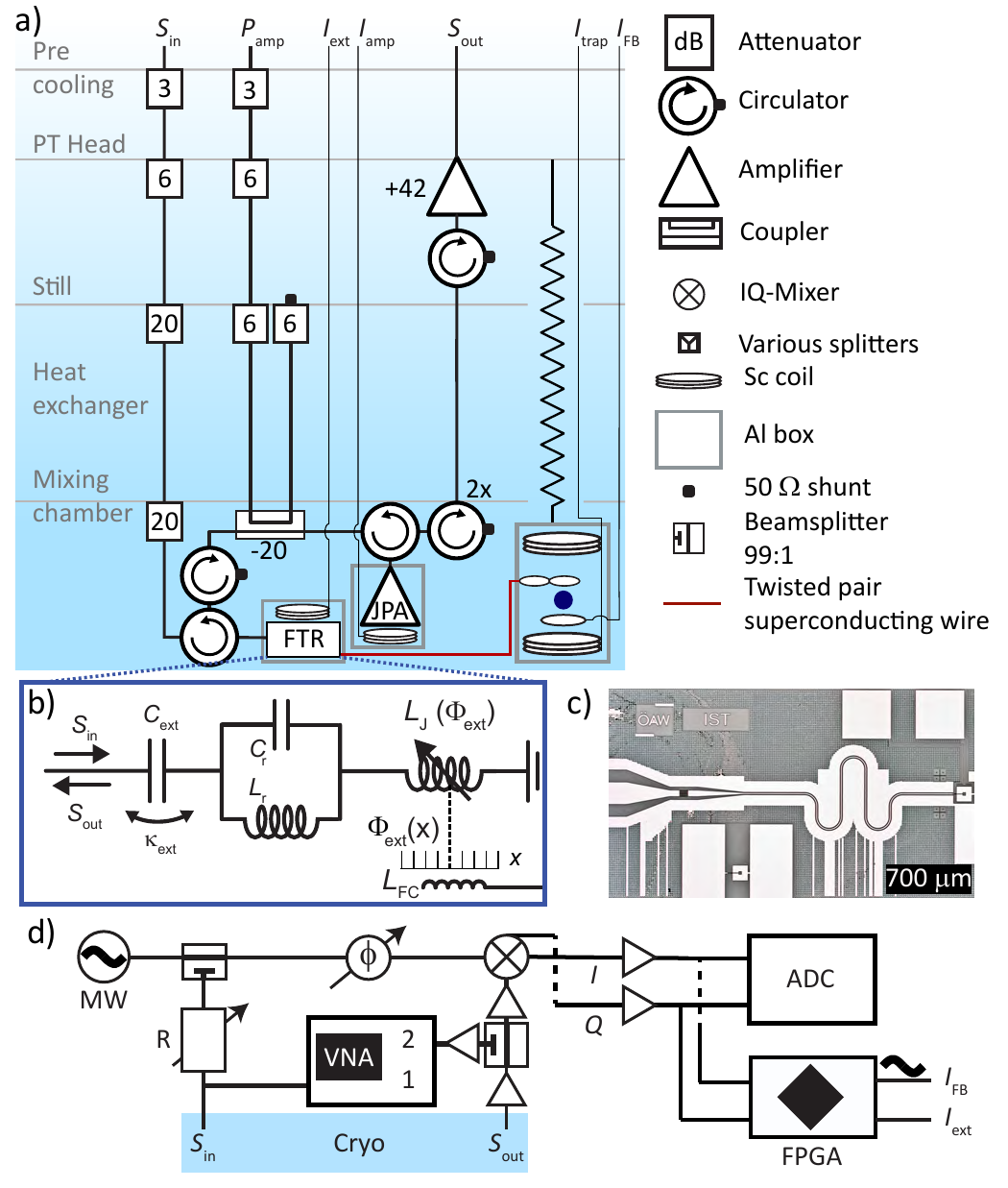}
        \caption{Panel a) The experimental platform is based on a levitated superconducting sphere coupled inductively by a gradiometric PUL to a flux-tunable resonator. The levitation platform is suspended from the 4K stage for vibration isolation and shielded by an aluminum box (grey) from magnetic fields. Two superconducting coils in anti-Helmholtz configuration with anti-parallel currents generate a field gradient to levitate the particle. A small coil allows to feedback cool the particle and acts as an input for a calibration tone. The particle's signal is guided to the cavity using a twisted pair superconducting wire. The microwave signal is screened from thermal radiation using in total \SI{60}{\dB} of attenuation. The input signal is reflected at the cavity and amplified using a high electron mobility amplifier at the $\SI{4}{\kelvin}$ stage. A Josephson parametric amplifier (JPA) was installed, but did not produce gain at the employed frequencies. Panel b) depicts the microchip consisting of a linear coplanar waveguide LC-oscillator and a SQUID shunting it to ground. Panel c) shows a microwave circuit of a sister sample. Panel d) sketches the room temperature microwave setup, having a microwave signal generator and a vector network analyzer as sources. The signals are detected either via the VNA or a digitizer. A field programmable gate array is used for cavity locking and to apply a flux calibration tone.}
        \label{fig:FullSetup}
\end{figure}

A schematic draft of the setup is shown in Fig.~\ref{fig:FullSetup}. Its parameters are summarized in Table \ref{tab:SampleParameters},  separated by the individual components consisting of: (i) the coplanar waveguide,  flux tunable, resonator (CPW-FTR), ii) the SQUID loop in a washer geometry (SQUID), (iii)  the flux transformator for remote sensing (flux trans.), and iv) the trap of the levitated superconductor (trap).

The mechanical oscillator consists of a superconducting sphere levitated in a magnetic field gradient. This trap has been detailed in Ref.~\cite{Hofer2023}. The trap gradient $b_i$ is controlled via $\Itrap$, and a calibration coil in close proximity to the particle is used for calibration of the induced flux in the microwave cavity via $\Ifb$.
The flux-tunable cavity consists of a distributed coplanar waveguide resonator. It is capacitively coupled via $\Cext$ to the signal line with an external coupling rate of $\kext / 2\pi \approx \SI{20}{\mega \hertz}$ and shunted to ground to form a $\lambda /4$ resonator with a SQUID at the end. The SQUID consists of a washer-type loop with inner diameter $\Ds = \SI{25}{\micro\metre}$ and outer diameter $\Ws = \SI{125}{\micro\metre}$, thus effective area of $\Aeff = (\SI{56}{\micro\metre})^2$ and geometric loop inductance $\Ls = \SI{0.12}{\nano\henry}$, derived from Refs.~\cite{Clarke2004, Gross1990}. The loop contains two Josephson junctions having a designed area size of $\AJJ = \SI{0.24}{\square\micro\metre}$ and a critical current of about $\Ics \approx \SI{0.5}{\micro\ampere}$ per single junction.
The applied microwave signal $\Sin$ is attenuated by $\Ltot = \Lrt \Latt \Lc \approx \SI{102}{\dB}$, composed of the microwave attenuation outside the cryostat $\Lrt = \SI{36.8}{\dB}$, the sum of attenuators $\Latt = \SI{49}{\dB}$, and the attenuation of the cables inside the cryostat $\Lc \approx \SI[separate-uncertainty=true]{16(2)}{\dB}$, set by the frequency dependent microwave transmission. The signal from a second microwave generator (MW) was additionally attenuated by $\SI{27.6}{\dB}$ and a switchable attenuator before joining the vector network analyzer (VNA) path.
After the attenuation the signal is reflected at the resonator and guided via circulators and a directional coupler to a Josephson parametric amplifier. With the parametric amplifier undriven in this work, the signal is simply reflected and sent to the HEMT, where it is amplified and reaches the room temperature part of the setup.

\subsection{Room temperature electronics}
The room temperature setup forms a homodyne interferometer in the microwave domain using an IQ-mixer for down-conversion. It is detailed in Fig.~\ref{fig:FullSetup}d. In particular, a microwave signal generator MW acts as photon source. The majority of the signal ($99\%$) enters the local oscillator (LO) arm that is matched in length with the signal path as the LO input of the mixer, which requires a strong and fixed power bias. An additional phase shifter ($\varphi$) enables fine tuning between the incoming signals at the IQ-mixer to rotate the I and Q quadratures if needed. The signal arm has a switchable attenuator (R), which we use to set the probing photon number. A splitter couples incoming tones from a VNA in. The signal is then sent to the input line of the cryostat $\Sin$. The outcoming signal from the cryostat $\Sout$ is amplified and split 99:1 with the minority being sent to the VNA for a direct measurement of the microwave cavity. The majority is further amplified and downconverted in the IQ-mixer. The two quadratures (I and Q) are amplified and split equally between an analog digital converter (ADC) and field programmable gate array (FPGA) used for cavity locking to ensure a constant flux sensitivity of it. The lock consisted of two proportional integral derivative (PID) controllers, where one locked slow, strong drifts ($\approx \SI{1}{\hertz}$) and a faster one that we regulated up to ($\approx \SI{40}{\hertz}$), such that the mechanical signal remained in the measured signal (detected at the ADC). The lock signals were sent to the bias coil of the FTR (slow) and the feedback signal of the trap (fast), see Fig. \ref{fig:FullSetup}a. In addition, a calibration tone was sent to the calibration coil to determine the flux ($\Sphiphi$) hence the frequency fluctuations ($\Sww$) of the microwave cavity.
 
\section{Details on calibration / from time domain to spectral densities}
\subsection{Motion}
As detailed in Ref.~\cite{Hofer2023}, we record the motion of the microsphere as a shadow image with a frame rate of $\SI{300}{\hertz}$ (or above). The frames are analyzed in terms of the position, sphere radius, and relative position to the PUL in pixels. Next the position data is converted into meters using the sphere's radius of $\SI{50}{\micro\metre}$, returning $\textbf{x}(t)$. We convert the time trace into a power spectrum (PS) using the Welch-Algorithm with a Hamming window and an effective noise bandwidth (ENBW) of around $\SI{0.09}{\hertz}$. The PS allows us to extract the mechanical frequencies (cf. Fig.~\ref{fig:Schematics}e) of each mode and its corresponding variance of the mechanical displacement ($\langle x_i \rangle^2  = $ PS($\Omega_i$)). 
\subsection{Flux calibration}
\label{sec:calibofcalibcoil}
For the flux calibration of the calibration coil we first determine the voltage to flux ratio of the external bias coil ($\Vext$), as picked up by the readout circuit. We find a flux periodicity of $\SI[separate-uncertainty=true]{467\pm1}{\milli\volt\per\Phi_0}$. We then flux lock the microwave cavity, sweep a static voltage at the calibration coil while recording the DC-shift of the flux lock through the external bias coil. We find a relation of $\SI[separate-uncertainty=true]{21.7\pm0.2}{\milli\volt}$ of countering voltage on $\Vext$ per applied Volt on the calibration coil. In combination with the calibration of the external coil we determine an induced flux of $\SI[separate-uncertainty=true]{46.5\pm0.4}{\milli\Phi_0}$ per applied Volt on $\Vfb$.

\subsection{Frequency fluctuations}
The microwave cavity is probed on resonance, homodyne downconverted using an IQ-mixer, fed into a digitizer, and recorded with a bandwidth of $\SI{1}{\kilo\hertz}$. For this mixing process we have matched the LO and rf paths of the microwave interferometer, however a finite rotation of I and Q depending on the probe frequency remained. To ensure to record the phase quadrature (thus the mechanical motion), we apply a quasi heterodyne mixing in post processing to the I and Q channels (at $\SI{400}{\hertz}$). The unwrapped phase of this mixing then contains the full data of the mechanical signal and the calibration tone of the calibration coil. Similar to the video data, we transform the recorded voltage trace via Welch's method with a 'Hamming' window of $1/10$ of the full data length into a power spectral density (PSD in $\mathrm{V}^2 / \mathrm{Hz}$) and PS in $\mathrm{V}^2$. The ENBW typically becomes $\enbw = \SI{114}{\milli\hertz}$ for a recording of $\SI{120}{\second}$, where $\mathrm{PSD} = \mathrm{PS} / \enbw$ and for an oscillation of $x(t)$ it holds that $\langle x_i(t) \rangle^2 = \mathrm{PS}(\Omega_i)$ \cite{Heinzel2002}.
We extract the peak heights of the PS trace for the individual peaks at $\Omega_i$ ($i$ in $x$, $y$, $z$ motion and calibration coil), corresponding to the detected mean value $\langle V_i \rangle^2 = \mathrm{PS}(\Omega_i)$ induced by a shift of the cavity frequency. 
The sphere's displacement was determined prior to and after the measurements via the recorded video, setting $\langle x_i(t) \rangle^2$. Typically, the amplitude fluctuations between the measurements were on the order of $25\, \%$. The calibrated displacement enables to shift the detected voltage to a displacement spectrum and via the ENBW to the displacement spectral noise density $\Sxx$, cf. left axis in Fig.~\ref{fig:DisplacementExample}. We note that, as $\langle x_i(t) \rangle^2$ differs for each mode, the calibration is repeated for every mode individually. Similarly, the calibrated flux noise density $\Sphiphi$ and the frequency fluctuations $\Sww = \Sphiphi (2\pi \slope)^2$ are derived, right axis in Fig.~\ref{fig:DisplacementExample}, by the calibration peak. The electromechanical coupling then relates to the induced frequency shift per motion with as: $G_i^2 = \Sww (\Omega_i) / \Sxx(\Omega_i)$.The measurement noise floor remained flat over the recorded frequency range, besides vibrations below our damping system (typically $<\SI{30}{\hertz}$) and multiples of the power grid oscillations $n\cdot\SI{50}{\hertz}$. We extract the imprecision noise floor from an off-resonance measurement at $\SI{475}{\hertz}$.

\section{Detector characterization}
\label{sec:Detector}
\begin{figure}
  \includegraphics[scale=1]{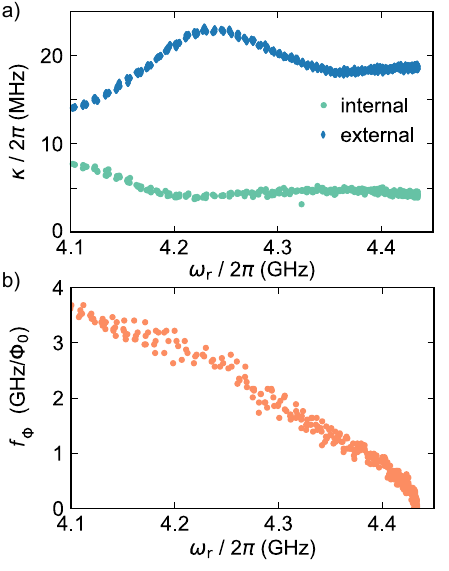}
        \caption{Details on the flux-tunable microwave cavity, extracted from the Fig.~\ref{fig:Schematics}b in the main text. Panel a) depicts the internal and external coupling rates. Over the full range, the cavity remained overcoupled with a total linewidth around $\SI{20}{\mega\hertz}$. Panel b) shows the frequency tuning as a function of bias frequency.}
        \label{fig:FTR_details}
\end{figure}

From the microwave spectroscopy measurements via the VNA of the bare flux tunable resonator, shown as blue datapoints in Fig.~\ref{fig:Schematics}b, c, and d, we fit the complex scattering parameter $S_{21}$ as \cite{Yamamoto2016, Pogorzalek2017}:
\begin{equation}
S_{21} = \frac{(\omega-\wres)^2 + i\kint(\omega-\wres) + (\kext^2 - \kint^2)/4}{\left((\omega-\wres) + i(\kext+ \kint)/2 \right)^2}
\end{equation}
(solid line in b, c), to obtain the resonator's eigenfrequency $\wres$ as well as internal $\kint$ and external linewidth $\kext$. The results of the loss rates are shown in Fig.~\ref{fig:FTR_details}a. We find a total linewidth of up to $\ktot / 2\pi = (\kint + \kext)/ 2\pi = \SI{30}{\mega\hertz}$, dominated by the external coupling over the full frequency range. The internal loss increased slightly for a resonator frequency below $\wres/2\pi = \SI{4.2}{\giga\hertz}$ but remained flat elsewhere. We note that the linewidth may be broadened by flux noise leading to an overestimate of the loss rate $\ktot$. We plan to improve the resonator loss mechanism to reach higher coherence, beneficial for the electromechanical cooperativity. Flux tunable resonators with internal losses around $\SI{2}{\mega\hertz}$ have been realized in nano-electromechanics and recent advancements offer promising prospects along this route down to expected $\SI{10}{\kilo\hertz}$ \cite{Schmidt2020, Chang2023}.\\
We note that upon connecting the trap, the linewidth of the cavity increased.

Moreover, we extracted the frequency tuning $\partial \wres / \partial \Phi = 2\pi \cdot \slope$ from the fits and depict the result in Fig~\ref{fig:FTR_details}b. The tunability $\slope$ is important to calibrate the flux sensitivity of the device and to tune the electromechanical coupling as $\gNull \propto \slope$. For the presented device we find a tuning of up to $\approx \SI{4}{\giga\hertz}/\Phi_0$ in the depicted range until $\SI{4.1}{\giga\hertz}$. We found the trend to continue at lower frequencies, but were limited by the working range of our cryogenic HEMT amplifier. Thus we consider tuneabilities of $\slope = \SI{10}{\giga \hertz}/\Phi_0$ easily in reach by simply increasing the resonator eigenfrequency, similar to demonstrated electromechanical circuits in Ref.~\cite{Schmidt2020}.

\begin{figure}
  \includegraphics[scale=1]{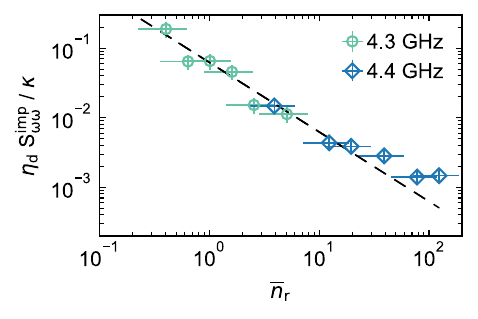}
        \caption{Frequency noise spectral density of the microwave cavity as function of the probe tone power. Notably, a linear reduction in noise is evident at both bias points. For higher read-out powers ($\approx 20$) we find the noise to decrease less than $1/\nr$, which we attribute to the inherent nonlinearity within the system. The data were taken at two different working spots of the cavity at $4.3$ and $\SI{4.4}{\giga\hertz}$ respectively. The dashed line shows a data trend with $1/\nr$.}
        \label{fig:cavnoise}
\end{figure}

Another important aspect of the electric resonator under scrutiny is its inherent nonlinearity. In contrast to linear cavity opto- or electromechanical systems, the radiation pressure cannot be increased arbitrarily by the number of read-out photons. Instead, a point is reached where the enhancement begins to diminish  \cite{Pirkkalainen2015}. We systematically assessed the imprecision noise levels as a function of the read-out photons, specifically targeting two distinct bias points of the cavity, each corresponding to differing coupling strengths. 
In order to facilitate a meaningful comparison between these datasets we have to take the different coupling strength, linewidth, and detection efficiency into account. We do this by re-scaling the imprecision noise as $\Simp = (16 \etad G^2 )/(\nr \kappa)$, with $\Sww^{\mathrm{imp}} = \Simp G^2$ and thus visualize $\etad^{\mathrm{cav}}  \Sww^{\mathrm{imp}}/\kappa$ as result in Fig. \ref{fig:cavnoise}. We find the cavity's noise floor to decrease with $\nr$ for both bias points as expected in the linear regime. However, for high photon numbers the decrease is getting less, which we attribute as onset of the cavity's nonlinearity at around $\nr \approx 20$.

\section{Electromechanical conversion}
\label{sec:Elmechconversion}

\subsection{Pick-up loop efficiency}
The induced flux of the particle is transferred to the SQUID with an efficiency $\atech = \Phiind / \Phim = \Mi / (\LPUL + \Lfc + \Ltw)$. The individual coil inductances are $\LPUL = \SI{0.9}{\nano\henry}$, $\Lfc \approx \SI{20}{\nano\henry}$, and $\Ltw \approx \SI{100}{\nano\henry}$. The dominant inductance thus is the transfer inductance between the coils, which also has the highest uncertainty as it is based on precharacterizations but may change due to the mounting geometry \cite{Hofer2023}. The mutual inductance of the SQUID and the input coil depends on the alignment of the flip chip, quantified by the input coupling coefficient $\nui$ as $\Mi = \nui \sqrt{\Lfc \Ls}$ with the SQUID loop inductance of $\Ls = \SI{0.12}{\nano\henry}$. We estimated the alignment at room temperature and believe drifts when cooling down to be negligible, in comparison to the dimension of the coil with a diameter of $\SI{200}{\micro\metre}$. For the alignment we reached, a numerical simulation yields a field factor of $\nui \approx 0.1$, hence we find $\Mi \approx \SI{160}{\pico\henry}$. Overall the flux transfer efficiency is estimated to be $\atech = 1.3\cdot10^{-3}$.

\subsection{Dependence of sensitivities on pickup position} \label{app:pickup_position}
We have determined a flux sensitivity of each mode to be $\partial \Phi / (\partial i) = (70, 800, 80)\,\Phi_0/\mathrm{m}$ at a trap gradient vector of $\bi = (39, 40, 80)\,\mathrm{T}/\mathrm{m}$. As the induced flux of the particle's motion scales with the applied gradient, the z-mode is expected to couple two times more strongly than the horizontal modes. However, we find that the sensitivity to particle motion along the y-direction is around an order of magnitude higher than the sensitivities to the particle motion in the other modes. This behavior arises from different sensitivities of the PUL to the different modes described by $\Ff_i$. This sensitivity depends on the PUL to particle displacement and the weak sensitivity to the z-mode indicates that the relative displacement of the PUL and the trap centre is such that $\Ff_z$ is close to a minimum.\\
\begin{figure}
  \includegraphics[scale=0.5]{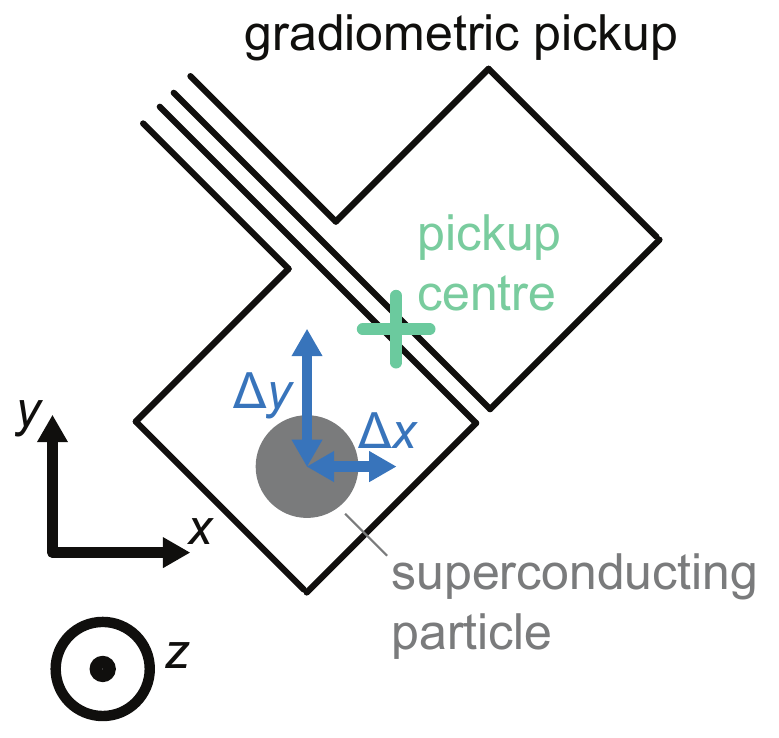}
        \caption{Schematic of the gradiometric PUL used in the experiment. It is rotated by $\SI{45}{\degree}$ with respect to the trap axes which enables  optical access to both $x$ and $y$ modes. The trap centre is displaced from the pickup centre by $\Delta x$ and $\Delta y$ as indicated, and also along the z direction by $\Delta z$. The particle is located at an arbitrary place in the schematic.}
        \label{fig:PUL_schematics}
\end{figure}
In this section, we describe how we used the experimentally determined flux coupling to estimate the separation between the pickup and the trap center. A schematic overview of experimental design is found in Fig.~\ref{fig:PUL_schematics}.\\
The flux sensitivity is calculated as described in Refs. \cite{Hofer2019, Hofer2023}: for a superconducting particle inside a quadrupole magnetic trap, the magnetic field at a point in space $\mathbf{r} = (x, y, z)$  can be described in terms of a vector potential $\magvecpot$, where $\mathbf{r_0} = (x_0, y_0, z_0)$ denotes a displacement of the superconducting particle from the centre of the quadrupole trap. From this  vector potential, we can derive the flux sensitivity to a particle displacement in the $r_0^i$ direction (with $r_0^i$ = $x_0$, $y_0$, $z_0$),  as picked up by a pickup loop, as:
\begin{equation}
\frac{\partial \Phi}{\partial r_0^i} = \int_{\mathcal L'} \frac{\partial \magvecpot}{\partial r_0^i} \cdot \mathrm{d}\mathbf{l'}, 
    \label{eq:dphidx_integral}
\end{equation}
where the integral is taken along the closed path $\mathcal L'$ of a pickup loop. We describe the gradiometric pickup loop as  two squares with side lengths $\SI{150}{\micro\metre}$ separated by a distance of $\SI{8}{\micro\metre}$, taking the path integral along opposite directions for the two squares. From this we obtain cumbersome analytical expressions for the flux sensitivity, which can be used to model the behavior of the flux coupling as a function of the pickup loop placement $\Delta \mathbf{r} = (\Delta x , \Delta y, \Delta z)$.
\\
\begin{figure}
  \includegraphics[scale=1]{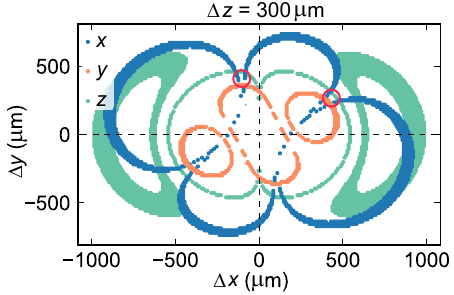}
        \caption{The relative flux sensitivities for a $z$-spacing of $\SI{300}{\micro\meter}$. The points of suitable sensitivities for the measured values are highlighted in the respective mode colors used in the main text. We find four possible particle positions, two highlighted in red, where the requirements for the mode sensitivities match within the particle size (circle radius = $\rp$). The two intersections in the lower part are point symmetric solutions to the two in the upper part (red circles).}
        \label{fig:placementxy}
\end{figure}
Upon assembling the setup with its geometry, the actual flux coupling remains undefined for four alignment parameters, that is the i) flux transformation $\atech$ estimated to be $1.3\cdot 10^{-3}$ given by the inductances in the setup, ii) the z-distance $\Delta z$ to the PUL, which we measure to be around $\SI{300}{\micro\metre}$ using the optical access, and iii) + iv) the distance in horizontal direction $\Delta x$, $\Delta y$, which we cannot measure when the experiment is set up. To reduce the complexity of the problem we thus first look at the relative sensitivity between the modes at a z-position of $\SI{300}{\micro\metre}$ as $\Delta x$ and $\Delta y$ are varied. The points of matching sensitivities with the experimental results are shown in Fig.~\ref{fig:placementxy} for each mode respectively. We find four positions where the sensitivity of all three modes agree within the particle size, highlighted with red circles. Mathematically speaking the four solutions correspond to two and a point symmetry along the PUL center such that intersection points in the top half are equivalent to the ones in the bottom half\\
\begin{figure}
  \includegraphics[scale=1]{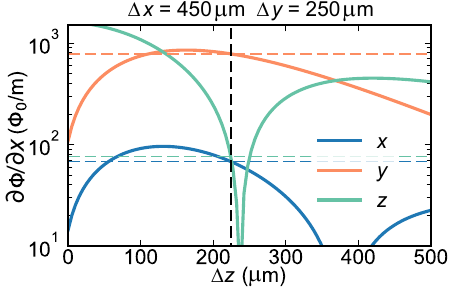}
        \caption{The flux coupling as a function of $z$-displacement as calculated from the presented model. The measured sensitivities are  represented by dashed lines. A PUL - trap center separation of about $\SI{250}{\micro\meter}$ describes the sensitivity best for $\Delta x = \SI{450}{\micro\metre}$ and $\Delta y = \SI{250}{\micro\metre}$.}
        \label{fig:sensitivities}
\end{figure}
Next, we determine the absolute flux coupling $\partial \Phi / \partial x$ at the intersection points as a function of $z$. We find best agreement for a flux transformation of $\atech = 5\cdot 10^{-3}$ and a PUL-particle separation of $(450, 250, 250)\,\mu \mathrm{m}$ as stated in the main text. This corresponds to the position in the top right quadrant of Fig.~\ref{fig:placementxy} and as expected in a coupling valley of the $z$-mode.\\
From Fig.~\ref{fig:sensitivities} one can also relate that a small shift in the particle's rest position affects the coupling significantly by the impact of the coupling valley. With the camera recording we measure $x(t)$ of the particle and thus can also measure the rest position of the particle with respect to the trap and PUL. When sweeping the trap current, hence $\bi$, we found the rest position to shift by $\Delta r = \sqrt{(\Delta x)^2 + (\Delta y)^2} > \rp$, larger than the particle radius (n.b. $\Delta \mathbf{r} \neq \Delta r$). We attribute this shift to minor alignment issues in the trap geometry. These shifts cause a large change in the flux sensitivity and thus without a precise value of $\Delta x$ and $\Delta y$ the modeling of $\gNull$ is challenging at this coupling position. Given the optical access of our experiment, we presently do not have access to it (but only to $\Delta r$).\\
The employed model of the flux coupling was used to calculate the improvements of the coupling at optimal positioning, which we find to be $(3.7, 0.3, 1.3)\cdot10^{6}$. However, these maximum points require very precise alignments, hence we consider improvements of up to $2\cdot10^{5}$ for the $z$-mode with the present generation of PULs to be realistic.

\section{A cavity electromechanical system using microwave photons}
\label{sec:optomech_app}
\begin{figure}
  \includegraphics[scale=1]{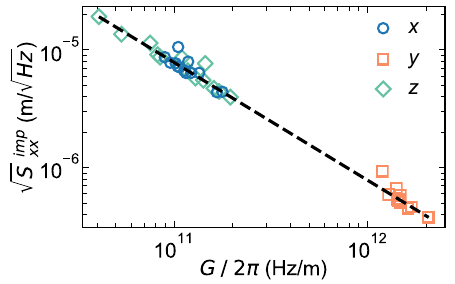}
        \caption{Performance of the electromechanical system. Each translational mode couples with a different strength due to the relative position of the sphere to the PUL. However, the imprecision noise density decrease is set by the cavity's properties and the detection efficiency (black dashed line). This efficiency is determined by the signal losses and the finite electronic noise temperature.}
        \label{fig:CavityOptomech_App}
\end{figure}
\subsection{Characterization of the detected imprecision noise}
We investigate the cavity electromechanical performance of the microwave cavity at a constant bias field at $\wres /2\pi= \SI{4.3}{\giga\hertz}$, corresponding to a working spot of $\slope = \SI{1.7}{\giga\hertz} / \Phi_0$. With the trap connected, the internal linewidth increases to $\kint /2\pi = \SI[separate-uncertainty=true]{110(13)}{\mega\hertz}$, the external one remained around $\kext /2\pi = \SI[separate-uncertainty=true]{25(5)}{\mega\hertz}$, thus $\ktot = \kint + \kext = \SI[separate-uncertainty=true]{135(19)}{\mega\hertz}$. We apply a weak probe tone corresponding to an average photon number of $\nr = 0.051 \pm 0.026$ in the microwave cavity, remaining deeply in the linear regime, as the cavity's nonlinearity typically becomes prominent above single-digit photon numbers ($\approx 10$) of similar cavities \cite{Schmidt2020}. The uncertainty in the photon numbers is dominated by the $\pm\SI{2}{\decibel}$ flatness in the microwave transmission. For a systematic analysis we sweep the trap current, by that affecting the coupling strength $G$ and the mechanical resonance frequency. As described in the main text we calibrate the measured voltage spectral density to a displacement spectral density and the cavity's frequency noise density. By that we obtain the electromechanical coupling strength as $G$ and $\gNull$, as well as the off-resonant imprecision noise floor $\Simp$, shown in a broader range in Fig.~\ref{fig:CavityOptomech} and zoomed in in Fig.~\ref{fig:CavityOptomech_App}. We find a decrease of $\Simp \cdot (G/2\pi)^2 = \SI[separate-uncertainty = true]{0.61(2)}{\tera\hertz}$ at this particular cavity bias condition (black dashed line).\\
For a noiseless, single-sided cavity electromechanical readout, probed on resonance, we expect the noise floor to scale as \cite{Aspelmeyer2014}
\begin{equation}
 \Sxximp (\omega) = \frac{\kappa}{16\nr G^2} \left(1+4\frac{\omega^2}{\kappa}\right),
\label{eq:SimpPerfect}
\end{equation}
where in the presented system $\omega \approx \Omegam \ll \kappa$, thus the second summand is negligible. Given the cavity bias parameters above, for perfect detection we find $\Sxximp (\omega) \cdot (G/2\pi)^2 = \SI[separate-uncertainty=true]{26(13)}{\mega\hertz}$. The increase in the detected imprecision noise $\Simp$ is related to the detection efficiency as $\etad \cdot \Simp = \Sxximp$, here $\etad = (4.3\pm2.1)\cdot10^{-5}$, when related to the fitted decay in the previous section \cite{Clerk2003}. The origin of this value is what we will discuss next.\\
We like to mention that the imprecision can be determined by experiment rather than the presented comparison to an ideal system \cite{Teufel2011}. However, this requires the back-action to become dominant, which was not the case for our realization.

\subsection{Characterization of the detection efficiency}
\label{sec:detefficiency}
\begin{figure*}
  \includegraphics[scale=1]{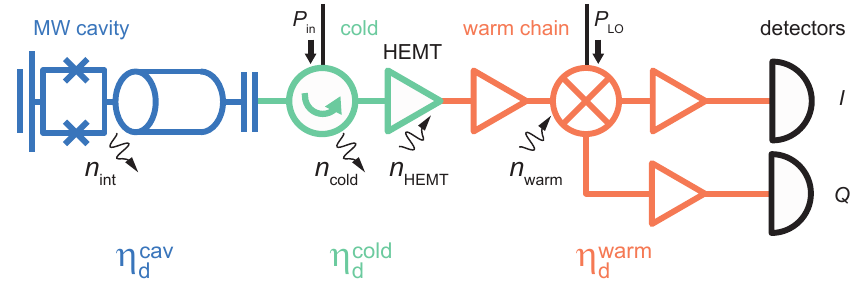}
        \caption{Schematics of the employed detection efficiencies. The total efficiency $\etad$ is distributed in the microwave cavity $\etacav$ (blue) attributed to internal photon losses, the efficiency up to the first amplifer $\etacryo$, also set by losses, and the added noise photons of the cryogenic HEMT amplifier (green). In addition, the remaining amplifier chain adds noise photons related to a detection efficiency $\etawarm$ (orange).}
        \label{fig:Detection efficiency}
\end{figure*}
We model the total detection efficiency to arise from the following three contributions, which are depicted in Fig.~\ref{fig:Detection efficiency}: The photon efficiency of the microwave cavity $\etacav$ (blue), the effective detection of the first employed amplifier $\etacryo$ (green), and the efficiency of the remaining amplifier chain $\etawarm$ (orange), thus $\etad = \etacav\cdot\etacryo\cdot\etawarm$. Any losses of the signal can be described by a beam-splitter adding a fraction of vacuum noise to the system. In Fig.~\ref{fig:Detection efficiency} we indicate losses as photons exiting the system and added noise by incoming photons.\\
One of the photon losses is described by the cavity efficiency, which relates how many photons leave to the environment as internal loss and not to the signal line. It is quantified by $\etacav = (1+\kint/\kext)^{-1} = 0.19\pm0.04$ at the particular cavity bias \cite{Schliesser2008}.
The noise from the remaining detection path relates to the efficiency of the amplifier cascade, which can be determined by Friis' formula as
\begin{equation}
 \Ttot = T_1 + \frac{T_2}{G_1} + \frac{T_3}{G_1 G_2} +\cdots,
    \label{eq:Frii}
\end{equation}
where each amplifier at position $i$ in the cascade contributes via its noise temperature $T_i$ and gain $G_i$. Losses beyond the first amplifier are considered negligible in comparison to the gain. Perfect detection $\etad = 1$ would require the amplifier chain temperature to be $\Ttot = 0$. Hence, the first amplifier having a low noise and high amplification is crucial for the overall performance. Most recent generations of quantum limited amplifiers and cryogenic HEMTs enable detection efficiencies of up to $\etacryo = 0.81$, given gain values of $\SI{30}{\decibel}$ and a HEMT performance of $\SI{1.5}{\kelvin}$. \cite{Winkel2020, Renger2021}. For this we employed a quantum limited Josephson parametric amplifier in our chain, cf. Fig.~\ref{fig:FullSetup}. However, for the final frequency range no significant gain was found and the HEMT amplifier became the first amplifier of the chain with a typical performance of $T_1 = \SI{2.5}{\kelvin}$ and $G_1 = \SI{42}{\decibel}$. In addition to this noise temperature relating to $n_{\mathrm{HEMT}}$ in Fig.~\ref{fig:Detection efficiency}, any loss on the way to the amplifier has to be taken into account ($n_{\mathrm{cold}}$). Besides careful design using superconducting microwave cables, the distance and used components between cavity and amplifier can cause significant losses, which are challenging to quantify in the cryostat. Therefore, we first want to address the accessible room temperature part of the amplifier chain.\\
It was designed to add less than $1\%$ of noise, relating to $\etawarm \approx 0.99$. However, by analyzing the dark noise of the amplifier chain we found an increase of the current noise by improper amplification of $\times 75$, thus $\etawarm = 1.3\cdot10^{-2}$. Fixing this efficiency enables an improved detection of the linear displacement spectral density of up to $8.7$ times from what is presented in the main text.\\
Since we have quantified $\etad = (4.3\pm2.1)\cdot10^{-5}$, $\etacav = 0.19\pm0.04$, and $\etawarm = 1.3\cdot10^{-2}$ the efficiency remaining for the cryogenic part between the coupling capacitance of the cavity and the input of the HEMT is $\etacryo = (1.8\pm0.9)\cdot10^{-2}$. This corresponds to $\nadd^{\mathrm{cryo}} = 28\pm14$, given $\etacryo =1/(1+2\nadd^{\mathrm{cryo}})$ \cite{Renger2021}. For a noise temperature of $\SI{2.5}{\kelvin}$, the HEMT amplifier adds $\nHEMT = 12$ noise photons. The discrepancy between the amplifier noise and the detected part arises from losses $\Lambda$ along the cabling. These can be treated as beamsplitters, enhancing the noise of the amplifier and adding vacuum noise as \cite{Teufel2011}
\begin{equation}
 \nadd^{\mathrm{cryo}} = \frac{\nHEMT}{\Lambda} + \frac{1-\Lambda}{\Lambda}\frac{1}{2}.
    \label{eq:CryoLosses}
\end{equation}
We find a loss of $\Lambda = \SI[separate-uncertainty = true]{-3.5(24)}{\decibel}$, which is on the typical order of loss in similar system \cite{Teufel2011}.\\
We summarize that the detection efficiency of the presented setup was determined to be $\etad = (4.3\pm2.1)\cdot10^{-5}$. A next generation with critically coupled cavity ($\etacav = 0.5$), a Josephson parametric quantum amplifier $\etacryo = 0.81$, and improved amplifier components at room temperature $\etawarm = 0.99$ bring a detection efficiency of $\etad = 0.4$ in reach, well above the required value of $0.11$ to resolve quantum mechanical states.

\subsection{Determination of the detector dark noise}
\begin{figure*}
  \includegraphics[scale=1]{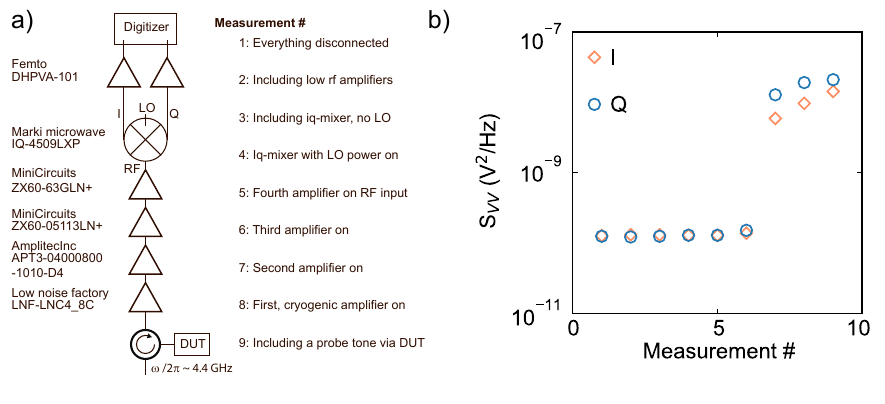}
        \caption{Measuring the voltage spectral noise density along the microwave amplification chain at the detector is essential for characterizing system performance. Panel a) provides a detailed description of the measurement setup and the sequence of measurements. Panel b) presents the measured noise floor density obtained from these measurements.}
        \label{fig:darknoise}
\end{figure*}

Upon conducting data analysis, we identified a significant discrepancy between the expected noise level of our amplification chain and the measured noise. To investigate the origin of this discrepancy, we performed a dark-noise measurement. Initially, all amplifiers and input signals were turned off. We conducted an initial measurement by unplugging the cables from our detector and performed a typical time trace measurement. We then determined the PSD $\SVV$, which yielded an average value of approximately $\SVV^{(1)}= \SI{1e-11}{\square\volt\per\hertz}$.\\
Subsequently, we systematically powered on the devices from the detector up to the device under test (DUT). Fig. \ref{fig:darknoise}a illustrates the sequence of measurements. The results are presented in Fig. \ref{fig:darknoise}b. The initial noise floor remained low until a notable increase in measurement 7. This observation suggests the source of the higher noise floor to stem from the third or second amplifier in the chain. Indeed, further investigation revealed that the third amplifier is not suitable for measurements within our frequency range, as it is specified to operate above 5 GHz rather than the applied 4.4 GHz. Comparing the final (full chain, no signal tone, measurement 8) and the initial noise level of the warm part (that is measurement 1), we find $\etawarm = \SVV^{(8)} / \SVV^{(1)} = 1.3\cdot10^{-2}$.

\subsection{Optimizing the system}
To highlight the capabilities of the device, we will briefly discuss potential future improvements of the presented platform. Key quantity for an optimized design to achieve ground state cooling is the quantum cooperativity, which can be written in terms of platform parameters as:
\begin{equation}
\Cq = \sqrt{\frac{3 \muN}{2 \rhop}} \frac{\BI \Itrap \nr \Qm \rp}{\kB \Tth \kappa} \left(2\pi \hbar \slope\alpha \beta \right)^2,
    \label{eq:CqDesign}
\end{equation}
where we introduced the temperature of the environmental bath surrounding the particle. In the optimization discussion, we will focus on electric design while keeping the trap and mechanical element unchanged. In particular, we keep the experimental $\BI, \Qm, \rp, \rho$ given in table \ref{tab:SampleParameters}, and assume the mode to thermalize at $\Tth = \SI{15}{\milli\kelvin}$, cf. Ref.~\cite{Hofer2023} for details. We will focus on the mode along the trap axis ($z$-direction). Potential improvements are discussed in the main text and the appendices and are summarized in the following:\\
In our measurements we found the cavity to decrease the imprecision level linearly with the readout power up to $\nr = 10$ photons, corresponding to an enhanced cooperativity of $200$ in comparison to the measurements in Fig.~\ref{fig:CavityOptomech}, cf. Appendix  \ref{sec:Detector}. We also note that recent work has investigated the benefits for cooling using the cavity's nonlinearity, which we have not considered in this discussion \cite{Bothner2022, Zoepfl2023}.\\
The most significant improvement in the setup can be achieved by enhancing the motion detection by positioning of the PUL to increase $\Ff$. Previous studies have already demonstrated the potential for achieving four orders of magnitude improvement \cite{Hofer2023, Gutierrez2023}. In our system, due to the larger displacement of the PUL compared to the Ref.~\cite{Hofer2023}, field calculations in Sec.~\ref{app:pickup_position} indicate feasible improvements of up to $2\cdot10^5$ (resulting in $\Cq\times4\cdot10^{10}$). Additionally, further enhancement can be realized by employing multi-loop gradiometers, as described in Ref.~\cite{Hofer2023}.
Guiding the flux to the SQUID loop is another important aspect that affects the system performance. This process is characterized by the scaling factor $\atech = \nui\sqrt{\Ls\Lfc}/(\LPUL + \Lfc + \Ltw)$, where $\nui$ represents the input coupling coefficient between the SQUID and the input coil. However, there is a high parasitic inductance between these two coils $\Ltw$, which hampers the flux guidance. Transitioning to a full-chip implementation reduces the parasitic inductance from approximately $\approx \SI{100}{\nano\henry}$ to $\SI{10}{\nano\henry}$. Additionally, improving the alignment when placing the flip-chip input on top of the SQUID loop can enhance the input coupling coefficient from the current $10\%$ to $50\%$ with a positioning precision of $\SI{10}{\micro\metre}$. Overall, these improvements in $\atech$ can boost the cooperativity by $2.5\cdot10^3$.
Furthermore, our measurements have been limited to flux sensitivities of $\slope = \SI{1.7}{\giga\hertz}/\Phi_0$, despite the detector's capability to reach $\slope = \SI{5}{\giga\hertz}/\Phi_0$, as determined in Sec.~\ref{sec:Detector}. Increasing the flux sensitivity to its maximum value improves the cooperativity by a factor of $10$. Recent advancements in flux-tunable resonators, with improved performance and narrower linewidths ($\kint /2\pi = \SI{40}{\kilo\hertz}$), offer promising prospects for further enhancing the sensitivity of our apparatus \cite{Chang2023}. For example, decreasing the linewidth from $\kappa /2\pi = \SI{30}{\mega\hertz}$ to a critically coupled cavity of $\SI{100}{\kilo\hertz}$ improves the cooperativity linearly, i.e. by a factor of $300$.
Additionally, incorporating a persistent current switch can reduce environmental noise, which is presently a limiting factor for the trap \cite{Hofer2023}. This switch also enables the use of high-power current sources, which can shift the particle frequency up to $\SI{650}{\hertz}$ and increase the coupling by a moderate factor of $2.1$, while simultaneously reducing the thermal excitation of the nanosphere. Consequently, this improvement enhances the cooperativity by a factor of $19$.
All these enhancements have the capacity to significantly elevate the low quantum cooperativity of $\Cq = 5\cdot10^{-17}$ in the presented measurement above $\Cq = 10^4$ for the next generation of the experiment, which is sufficiently high for the detection of single phonon quanta.

\bibliography{main}

\end{document}